\documentclass[twocolumn,amssymb, amsmath,nobibnotes, aps, prl,showpacs]{revtex4}
\usepackage[dvips]{graphicx}
\begin{document}
\title{Growth of large single crystals of Rare Earth Hexaborides}
\author {G.Balakrishnan, M.R. Lees and D.M$^{c}$K. Paul }
\affiliation{Department of Physics, University of Warwick, Coventry CV4 7AL,U.K}

\begin{abstract}

Single crystal growth of several rare earth hexaborides has been carried out by the floating zone technique. A high power Xenon arc lamp image furnace was used for the crystal growth. Large high quality crystals of LaB$_{6}$, CeB$_{6}$, PrB$_{6}$ and NdB$_{6}$, about 1 cc in volume have been obtained. Crystals of all these compounds have also been grown using enriched $^{11}$B isotope for use in neutron scattering experiments.

\end{abstract}
\maketitle
\section{Introduction}\label{1}Rare earth hexaborides are an interesting class of compounds. The alkaline earth hexaborides have been investigated recently for their extraordinary magnetic properties when doped with La, for instance, giving rise to ferromagnetic behaviour, with a high ordering temperature  and a low moment\cite{Young}. CeB$_{6}$, exhibits an extremely interesting and intriguing magnetic phase diagram at low temperatures with both magnetic and quadrupolar ordering\cite{Effantin}. In order to investigate the magnetic properties in detail using neutron scattering techniques, it is often necessary to conduct experiments on a large volume of the sample. Crystals of the hexaborides grown using Al flux are invariable too small for these measurements. The growth of large crystals of these materials has therefore only been possible by the floating zone technique. In the past this has often been done using R.F power, owing to the very highmelting point of these compounds\cite{Otani1}. Crystals of CaB$_{6}$ and SrB$_{6}$ are difficult to obtain directly from the melt by the floating zone  method due to their high volatility. LaB$_{6}$ has been grown successfully by the floating zone method previously\cite{Otani2} and is probably the best studied member of this class of compounds as regards its crystal growth owing toits extensive applications as an efficient electron emitter.  To date the  magnetic properties of the Nd and Pr hexaborides have not been extensively studied.Along with the La and Ce compounds, these materials have the advantage of having congruent melting points. This enables the growth of large single crystals of these compounds, without the use of a flux. The hexaborides have melting points around 2500 C, and this temperature varies depending on the boron content in the compounds\cite{Otani1}. An Image furnace equipped with Xenon arc lamps is capable of reaching temperatures of 2800 C and is therefore ideally suited for the crystal growth of these compounds. This paper describes the crystal growth of large, high quality single crystals of the Ce, La, Nd and Pr hexaborides by the floating zone method, using such a furnace.    

\section{Experimental Procedure}\label{2}Commercial rare earth hexaboride powders (Cerac, USA, 99.9\%) were used as starting materials.  For the crystal growth of compounds with enriched $^{11}$B isotope, powders of the hexaborides were synthesized starting from the respective rare earth oxides (RO$_{x}$) and the enriched $^{11}$B isotope powder. All the rare earth oxides were of high purity (99.9\%) and the $^{11}$B isotope powder was  99.5 \% enriched (Eagle Picher, USA). The powder mixtures containing  a slight excess of boron were mixed well and reacted in vacuum. The resulting powder was ground well and isostatically pressed to produce rods of 6 to 8 mm diameter and 70 mm length. In some cases it was necessary to add a binder such as PVA or PVB to the powder prior to pressing. The rods were then sintered in a flow of argon gas at 1550 C for 1 hour. Prior to argon flow during the sintering process, the furnace was evacuated to give a vacuum of $\sim$ 10$^{-5}$ mbar($\sim$ 10$^{-3}$ Pa). The sintered rods were then used for the crystal growth.Crystal growth was carried out by the floating zone method using a Xenon arc image furnace 0. This is a four mirror furnace equipped with 3kW  Xenon arc lamps and is capable of reaching 2800 C. The crystal growths were done in a flow of argon gas, 6-10 l/min, with the feed and seed rods rotating at 30 rpm. The growth speeds ranged from 10 to 18 mm/h. The crystal boules produced were first examined using X-ray Laue diffraction. Pieces for various measurements were cut from the boule by spark erosion. \section{Results and Discussion}\label{3 }The crystal growth of the Ce, La, Pr and Nd hexaborides is made easy by the fact that all these compounds have congruent melting points. The molten zone was very stable for the entire duration of the growth in all cases and the power needed to melt the rods was around 80-85\% of the total output power of the lamps.  All the boules obtained were  a deep purple to deep bluish purple colour. The as grown boules of the Ce, Nd and Pr hexaborides all had a fine coating of a white speckled crust that was easily removed by an abrasive material, to reveal a shiny surface. This coating was not seen in the LaB$_{6}$ crystals grown. The only difference in the preparation of the LaB$_{6}$ crystal was that there was no binder used to make the starting rods. The exact chemical composition of the white `crust' is being studied at the moment. The boules were roughly5 mm in diameter and about 50 to 60 mm long. There was no difference in the growth conditions for the crystals containing the Boron isotope. Figure 1 shows the photographs of the as grown boules of two of the rare earth hexaboride crystals. It is in principle, possible to grow crystals of larger diameter than the ones grown here, for most materials. Normally such a growth would require a higher power to melt the larger diameter starting  rods and in the case of the hexaborides the power required would then be close to the maximum power achievable using the Xenon arc lamps.

\begin{figure}
\includegraphics[width = 8.0 cm]{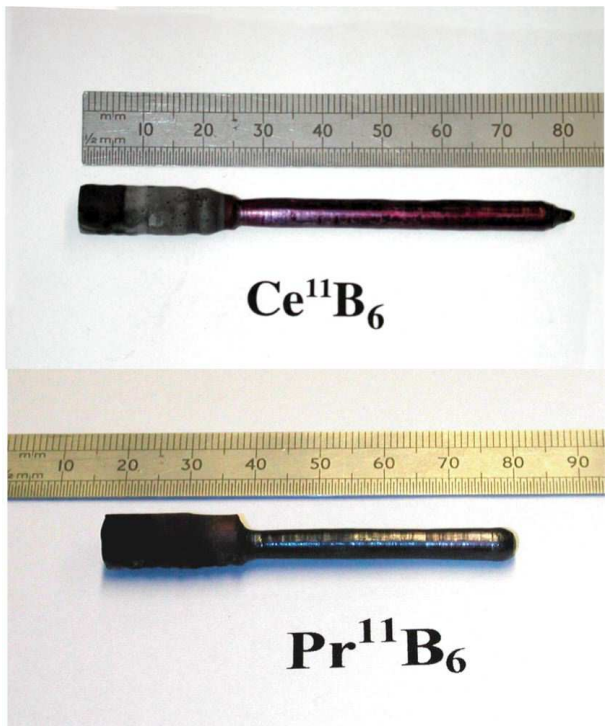}
\caption { Photograph of the as-grown boules of CeB$_{6}$ and PrB$_{6}$ grown by the floating zone technique. These crystals were grown using enriched $^{11}$B isotope for use in neutron scatteringexperiments.} 
\end{figure}

As the hexaborides are inherently hard materials\cite{Otani1}, they could not be cut using diamond coated blades ona slow speed saw. Spark erosion was used to cut slices from the grown boules for tests and various measurements.  The  crystals grown were examined using X-ray Laue diffraction to check their quality. Figure 2 shows the X-ray Laue diffraction photograph taken on a crystal of LaB$_{6}$. The crystals did not need to be seeded.  They picked up an orientation along one of the principal axes, often the (100),very quickly even with a polycrystalline seed rod. The first growth was carried out using a polycrystalline rod as a seed and the crystal grown was then used as a seed for subsequent growths. 

\begin{figure}
\centering
\includegraphics[width = 7 cm]{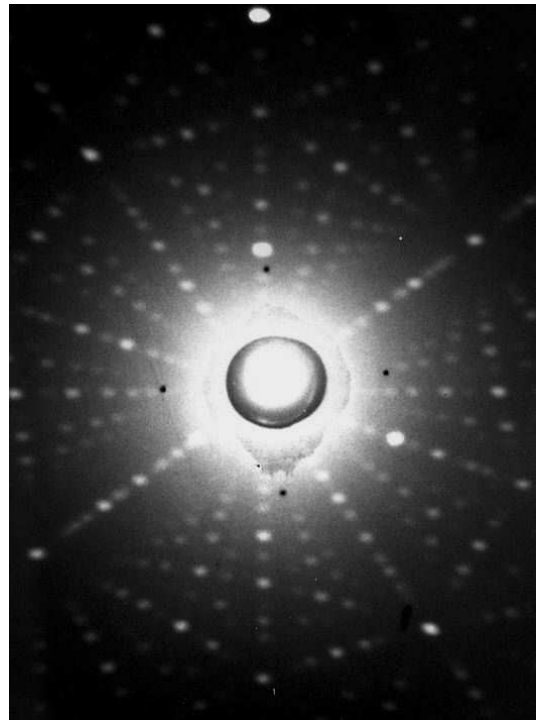}
\caption {X-ray Laue back reflection photograph of a crystal of LaB$_{6}$ showing the (100) orientation.}
\end{figure}

The size of the crystals that can be grown using the floating zone technique is much larger than those that can be obtained using conventional flux growth. As a comparison, we have also carried out the crystal growth of a few alkaline earth hexaborides using  Al as flux (0.2 at.\% of Ca/Sr, 99.8 at.\% Al). These crystals  are much smaller in size, the maximum sizes obtained for CaB$_{6}$ being 5 mm x 2 mm x $\sim$ 0.1 mm.  Figure 3 shows the photographs of crystals of CaB$_{6}$  and SrB$_{6}$ grown using Al flux. The CaB$_{6}$ crystals form as thin platelets and are fragile. SrB$_{6}$ crystals grown by the aluminium flux method, are also small and avariety of shapes and sizes are obtained. Most of them form as rods or needles while some form as platelets.An electron micrograph of a small section of a rod shaped crystal of SrB$_{6}$ is  shown in Figure 3. The platelets of SrB$_{6}$ showed traces of aluminium present on the surface when examined using an electron microscope. As expected, the crystals grown by the floating zone method were free of any contaminants. 

\begin{figure}
\centering
\includegraphics[width = 7.0 cm]{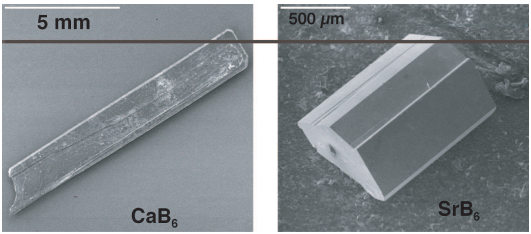}
\caption {CaB$_{6}$ crystal and a section of a SrB$_{6}$ crystal, both grown from the melt using aluminium flux shown here for comparison with those of the rare earth hexaborides.}
\end{figure}

\section{Summary and conclusions}\label{4}Single crystals of the rare earth hexaborides, RB$_{6}$ for R= La,Ce,Pr and Nd have been grown by the floating zone technique, using a four mirror Xenon arc Image Furnace. All these compounds melt congruently and therefore are easy to grow by this method. The  RB$_{6}$ crystals obtained by this method are much larger than those of CaB$_{6}$ and SrB$_{6}$ that were grown from the melt using Al as flux.Preliminary examination with X-rays indicates that the quality of the crystals grown is good. As no flux materials or crucibles are used in the RB$_{6}$ growth, the crystals obtained are free from any contaminants. Experiments toexamine the magnetic properties of these crystals are currently underway in the laboratory. Neutron scattering studies are planned on the magnetic phases using the large crystals obtained with the enriched $^{11}$B isotope. \section {Acknowledgement}This work was supported by a grant from the EPSRC, UK.We wish to thank Martin Davis and Steve York for their assistance in electron microscopy and  composition analysis of the crystals.

\end {document}